# NeuroStorm: Accelerating Brain Science Discovery in the Cloud

Gregory Kiar[1], Robert J. Anderson[2], Alex Baden[3], Alexandra Badea[2], Eric W. Bridgeford[3], Andrew Champion[7], Vikram Chandrashekhar[3], Forrest Collman[10], Brandon Duderstadt[3], Alan C. Evans[1], Florian Engert[9], Benjamin Falk[3], Tristan Glatard[4], William R. Gray Roncal[3], David N. Kennedy[3], Jeremy Maitin-Shepard[8], Ryan A. Marren[3], Onyeka Nnaemeka[9], Eric Perlman[3], Sharmishtaas Seshamani[10], Eric T. Trautman[7], Daniel J. Tward[3], Pedro Antonio Valdés-Sosa[5,6], Qing Wang[6], Michael I. Miller[3], Randal Burns[3], Joshua T. Vogelstein[3]

[1]McGill University, USA [2]Duke University, USA [3]Johns Hopkins University, USA [4]Concordia University, Canada [5]The Clinical Hospital of Chengdu Brain Science Institute, MOE Key Lab for Neuroinformation, University of Electronic Science and Technology of China, Chengdu 611731, Peoples Republic of China [6]Cuban Neuroscience Center, Cuba [7]Janelia Research Campus, USA [8]Google, USA [9]Harvard University, USA [10]Allen Institute For Brain Science, USA [11]University of Massachusetts Medical School, USA

Neuroscientists are now able to acquire data at staggering rates across spatiotemporal scales. However, our ability to capitalize on existing datasets, tools, and intellectual capacities is hampered by technical challenges. The key barriers to accelerating scientific discovery correspond to the FAIR data principles: findability, global access to data, software interoperability, and reproducibility/re-usability. We recently conducted a hackathon dedicated to making strides in those steps. This manuscript is a technical report summarizing these achievements, and we hope serves as an example of the effectiveness of focused, deliberate hackathons towards the advancement of our quickly-evolving field.

## 1 Introduction

Neuroscientists from around the world are building tools to process and disseminate next-generation datasets. These datasets range from nanometer to millimeter resolution, from millisecond to lifetime temporal resolution, and from C. elegans to humans. Collecting and analyzing these datasets can be difficult and costly. It is therefore advantageous to the community that we engage as many scientists (both professional and citizen) as possible.

To democratize neuroscience, enabling anybody to both generate and consume neuroscience knowledge, requires change. A series of workshops held over the last couple years discussed how to facilitate this change. The first workshop proposed a series of grand challenges that were maximally significant, feasible, and inclusive [1]. The group determined that successfully addressing each of these questions would require a common universal resource: a cloud platform [2]. The second workshop discussed these ideas with a larger community to get feedback, further refining these ideas [3]. The third workshop focused on determining the actionable next steps that we could achieve in a three month period. Finally, a hackathon took place before the annual Society for Neuroscience conference in October 2017, in which we took those steps. This manuscript describes those steps.

Each project we embarked upon adhered to at least one of the FAIR principles [4]: findable, accessible, interoperable, or re-usable/reproducible. More specifically, findability and global accessibility is achieved by making all software and data available in a searchable resource. Interoperability is achieved by utilizing existing APIs to write conversion layers between software tools. And reproducibility is achieved by using common standards.

The culmination of this work sets the stage for future hackathons and other unified, focused efforts towards accessible and impactful advances in brain sciences.

## 2  Global Democratization

**Zebrafish Light Microscopy Stacks in the Cloud**   Alongside the original macroscopic Zebrafish Brain (Z-Brain) data collected and stored with NeuroData [5], the newer multispectral Z-Brain volumes [6] (over 100 immunohistochemical stains) have been integrated with BossDB, significantly increasing the accessibility of the dataset. Scientists can leverage these datasets to perform analyses at a scale and resolution that was previously impossible in the zebrafish larval brain.

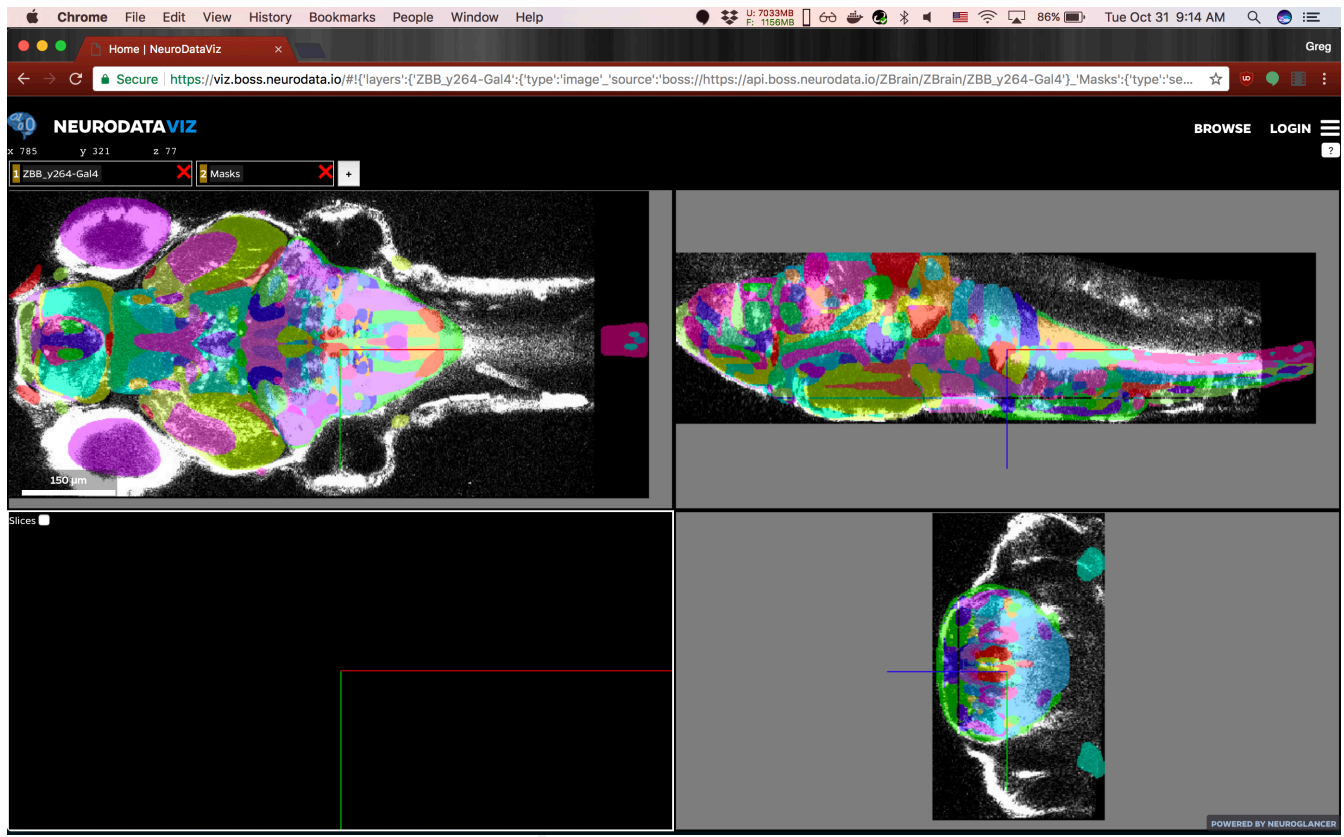

Figure 1: The Z-Brain dataset and Zebrafish atlas as viewed on the web using NeuroData's visualization engine from data stored in BossDB on the Amazon Web Services cloud.

**BigBrain in the Cloud**   The BigBrain [7] dataset has been ingested into Neurodata's public BossDB instance (Figure 2). This enables spatial querying and web-visualization of the 20 $\mu$m resolution histological image of the human brain. With this integration scientists can now expressly query specific regions of BigBrain, perform parallelized processing on portions of the total volume, and view the data in the cloud [1]. This increases the accessibility of BigBrain, and reduces the computational burden for experimenting with processing algorithms on this unique and valuable dataset.

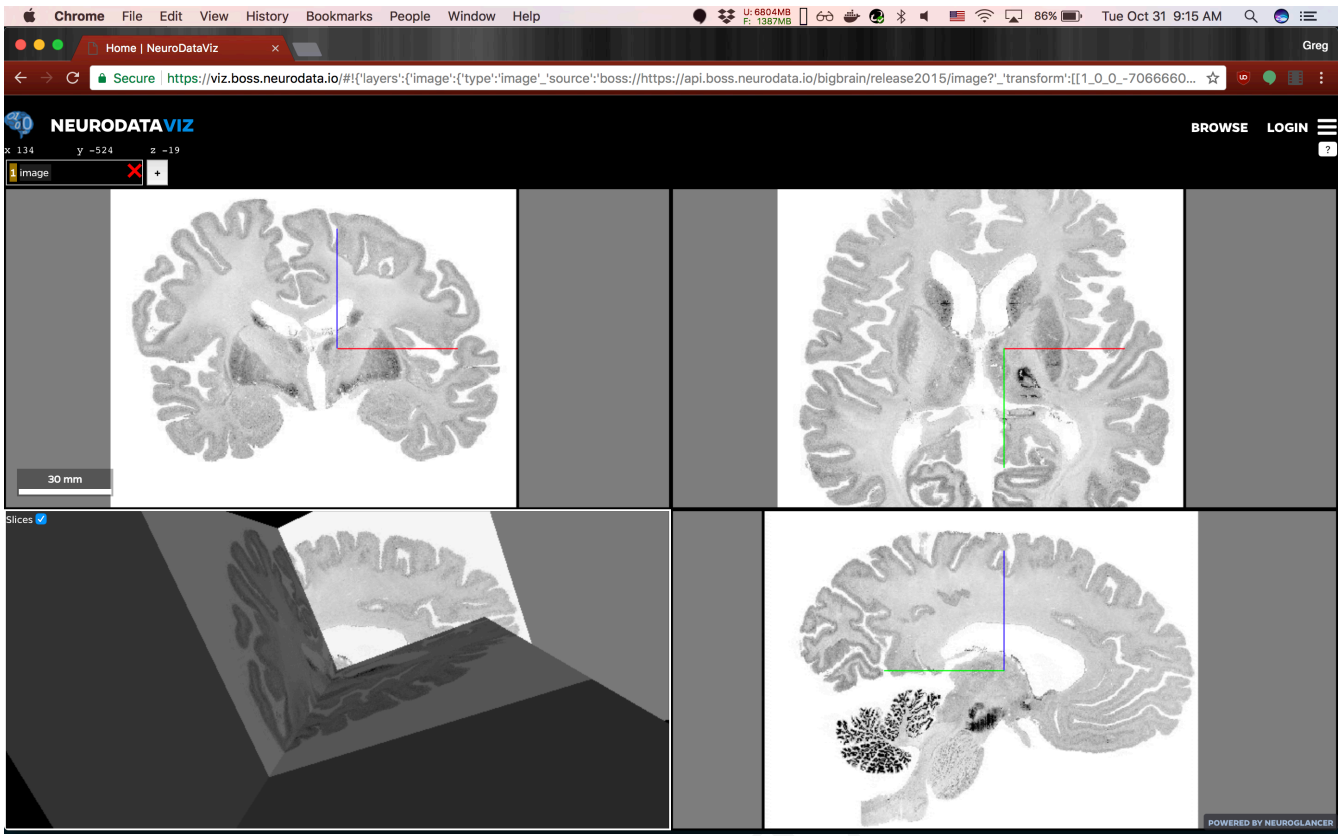

Figure 2: The BigBrain dataset as viewable on the web using NeuroData's visualization engine using data stored in BossDB on the Amazon Web Services cloud.

**High Performance Computing on the Chinese Cloud**   The web-based cloud resources management platform, CBRAIN, has been deployed and tested on the Baidu cloud in China. It is currently managed by the team from the Joint China-Cuba Lab for research in Translational Neurotechnology, with help from MNI, McGill. The computing system is based on a SLURM cluster and has been deployed on Baidu Cloud, which is one of the largest cloud companies in China. This integration extends easy-access to high performance computing and cloud based neuroimaging data sharing to Chinese and Cuban researchers.

**Latin American Cloud Neuroscience Environment**   On March $8^{th}$, 2008 in Havana, the Latin American Network for Brain Mapping (LABMAN) was created with participants from Argentina, Brazil, Colombia, Cuba, and Mexico. LABMAN aims to promote neuroimaging and systems neuroscience in Latin America, and to increase public awareness of the potential for Latin American scientists to contribute to both basic and applied research in human brain mapping. Networking status shows high heterogeneities within and between Latin American countries, thus country-specific solutions are needed. For example, relationships between the United States and Cuba are currently not feasible. In these cases, some Chinese companies or groups can serve as substitutes; alternatively, the C-C-C (China-Cuba-Canada) collaboration project can also provide cloud storage and computing power for the Cuban scientists.

**QUantitative Behavioral Assessment (QUBA)** Initial discussions occurred at NeuroStorm regarding an international open science project designed to lower the barriers and expense to facilitating a quantitative behavioral assessment in the context of the child psychiatry evaluation. This initial planning is documented at: `https://tinyurl.com/y7h3lw8f`. The output of these discussions is an initial plan to develop a minimally invasive, cost effective, fully automated quantitative behavioral and physiological assessment protocol to serve as an adjunct to a typical child psychiatry clinical visit. Conceptually, in addition to the subject of the assessment, this scheme consists of: A Room; and The Personnel Sensors. Armed with this Room and Wired Subject, within a 15 minute structured assessment protocol, a comprehensive quantitative report can be made of: language assessment, auditory evoked potential, anxiety reactivity, gait assessment, heart rate variability, eye tracking, etc.

With minimal effort and cost, these data can be augmented by microbiome and genetic assessment. Followup planning will occur to begin to develop clinical partners, industry relationships, initial technology development grant proposals, and continued community engagement in this concept.

**Neurodata's Opensource Method for Autonomous Detection of Synapses (NOMADS)** Current methods of synapse detection in Light Microscopy (LM) imaging data are difficult to access, require a prohibitively large amount of training data, or depend on critical user facing hyperparameters. These factors make it nearly impossible for neuroscientists around the world to utilize these tools. We developed NOMADS, an unsupervised, hyperparameterless method for detection of synapses in LM data. The NOMADS pipeline performs at or above the precision and recall of cutting edge LM synapse detection methods, and can be easily accessed by researchers around the globe through DockerHub[2] and Github[3].

## 2.1 Interoperability

**Image Database For Researchers in the Cloud** The BossDB platform was developed by the Johns Hopkins University Applied Physics Lab, and is used by NeuroData to store large image volumes. BossDB enables spatial queries, visualization, and data processing in the cloud, with many datasets publicly accessible. This platform has extended the usability of BOSS by creating a new client for data ingest to be used by researchers[4]. An open-source front-end for downloading data in BossDB and for generating links for visualization[5] was also created[6].

**Cloud Deployment of Tile Database** The render [8] framework for storing and displaying collections of transformed image tiles has been deployed on AWS, and a series of improvements that Docker-ize and parameterize components of this web service have been made. For example, support for render to read data from AWS S3 buckets was added, and bugs related to calculation of stack bounds were made. Migration scripts were also improved to facilitate the transfer of render stacks (combining image data and associated transforms) between deployments.

**Hierarchical Image Alignment Tool** A new alignment strategy being developed at Janelia Research Campus uses multiple alignments of montages rendered at different scales. The overall strategy was reviewed and a prototype stack partitioner was developed. The prototype was tested using the hackathon's AWS render deployment. Multiple minor improvements and bug fixes were made to facilitate usage of the AWS deployment for testing.

**Image Sharing Tools** CATMAID [9], a web-based annotation toolkit used on several large reconstruction projects, was extended to support BossDB as an image source. This can enable a user to do their own annotation within a publicly accessible BossDB data source. A new Neuroglancer data source for displaying CATMAID-drawn skeletons was written to allow co-registered display of volumetric image data and meshes (such as annotated surfaces), alongside the underlying image data. Support was also added for Neuroglancer to display Render stacks with multi-channel image tiles. Through Neuroglancer images can be visualized and dynamically explored in all three orthogonal planes, and will soon be able to integrate additional surface mesh models.

## 2.2 Reproducibility

**Integration of EEG into Web-Based High Performance Computing Environments** Electrophyisology provides functional data with high temporal resolution that is crucial to the study of brain functions and disorders, and serves as a viable translational bridge in all economic settings. In spite of this, it is an imaging modality neglected in recent Global Brain Projects [10]. Part of the mission of the C-C-C collaboration is to better integrate electrophysiology (EEG and MEG) into the CBRAIN platform, a web-based cluster management platform and portal [11]. Towards this goal, we have initiated efforts to: **a)** incorporate the tomographic quantitative EEG (qEEGt) toolbox [12] developed by CNEURO into CBRAIN via Docker and Boutiques; **b)** adopt the BIDS-EEG format for storing data into the LORIS [13], the data management platform of CBRAIN; **c)** transplant the visualization toolbox of EEG data to CBRAIN; and **d)** release, as part of the data repositories, the data from the Cuban Human Brain Mapping Project [14]. To our knowledge, this is the only national effort to integrate EEG with other imaging modalities.

**Integration of Standards for Data and Computational Pipelines** The Boutiques [15] descriptive command-line framework, supported by a Python package, has been extended to include a BIDS app [16] importer, that generates descriptors from BIDS applications. This enables infrastructures, such as CBRAIN [11] and the Virtual Imaging Platform [17], which support Boutiques, to natively support the execution of BIDS apps as well, further increasing the accessibility of BIDS datasets and supporting applications.

**Structural Connectome Estimation in the Cloud** The NDMG [18] structural and functional connectome estimation pipelines now both fully support the BIDS app [16] specification. These tools can be easily integrated with existing infrastructures that support the curation of BIDS [19] datasets, and platforms that execute BIDS applications. This provides the community an accessible entrypoint for generating reliable estimates of multimodal human brain connectivity and performing quality control of intermediate and terminal derivatives.

**Sub-Cortical Structure Shape Analysis** Tools for analyzing patterns of subcortical gray matter atrophy or growth in populations affected by aging and disease have been described in [20] and have previously been made available through MRICloud[7] [21]. These tools operate on triangulated surface meshes and are based on the large deformation diffeomorphic metric mapping [22] framework. They have been updated to accept input datasets in the standardized BIDS format [19], and have been packaged into a docker container as a BIDS app [16]. These resources will soon be available publicly through the BIDS Apps Forge[8].

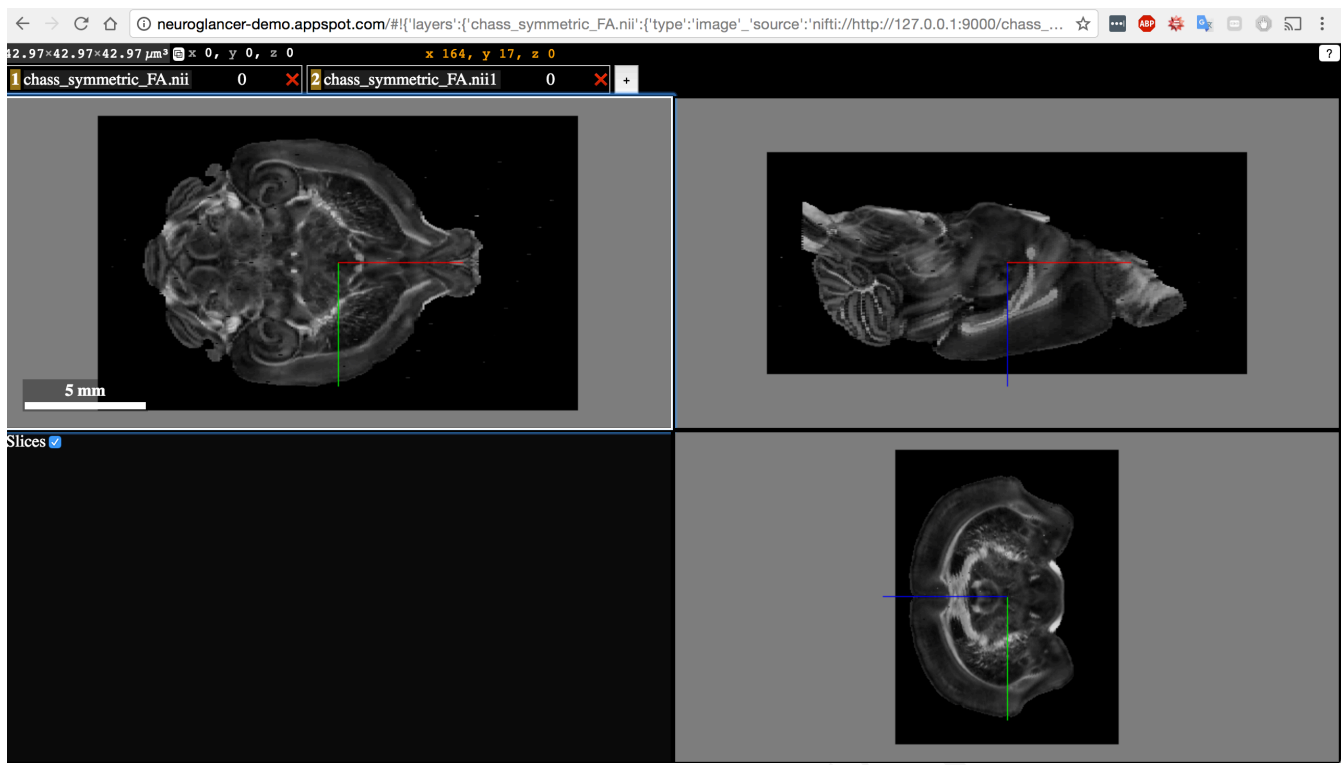

Figure 3: The mouse brain dataset as viewable using Neuroglancer

**Small Animal Multivariate Image Analysis (SAMBA) Pipelines**  A manuscript has been submitted for publication, describing a high performance computing cluster implementation of a pipeline for small animal brain image analysis. The pipeline produces regional and voxel-wise statistical analyses, and prepares data for connectomics. SAMBA is centered primarily on high spatial and angular resolution diffusion weighted MRI, and can be used with multiple image channels, providing multivariate biomarkers. Importantly it also introduces a comprehensive validation framework. The source code is also shared at https://github.com/andersonion/SAMBA. We established the concrete steps needed to adapt the pipelines to a portable format using docker. There are plans to modify the internal file structure to support the BIDS app [16] specification, and to also add Boutiques command-line support. We explored novel sharing and visualization possibilities (see Fig. 3) for small animal images, using Neuroglancer.

**Cleared Brain Registration Package**  An easy-to-use affine and LDDMM registration package for cleared brains [23] was developed and published on DockerHub[9] and Github[10]. One application demonstrating the registration of a CLARITY [24] image to the Allen Reference Atlas [25] and parcellation at 50um resolution is shown in Fig. 4.

**PANDA EEG Preprocessing Pipeline**  PANDA is a six step EEG denoising pipeline which takes as input raw EEG data and returns preprocessed EEG data along with quality assurance information and figures. The pipeline starts by centering each channel about its mean, then high pass filters each channel at 1 Hz. Next, a linear regression technique is used to remove eye movement artifacts detected from the EOG channels, a wavelet based denoising technique is used to remove high amplitude spikes due to muscle artifacts, and a shrinkage operator is applied to adjust the magnitude of time steps 8 or more standard deviations away from the mean of each channel. Finally, bad electrodes are deemed to be those with kurtosis or KS statistics 2 or more standard deviations away from the mean statistic for all channels, and are interpolated by averaging the wavelet coefficients of 3 nearby channels. The steps, ordering, and default hyper-parameters of the pipeline were chosen to maximize the probability that two scans from the same subject are have a smaller euclidean distance between their correlation matrices than two scans from different subjects. This pipeline was designed to be run locally or at scale on cloud computing infrastructures, and is published on DockerHub[11] and Github[12].

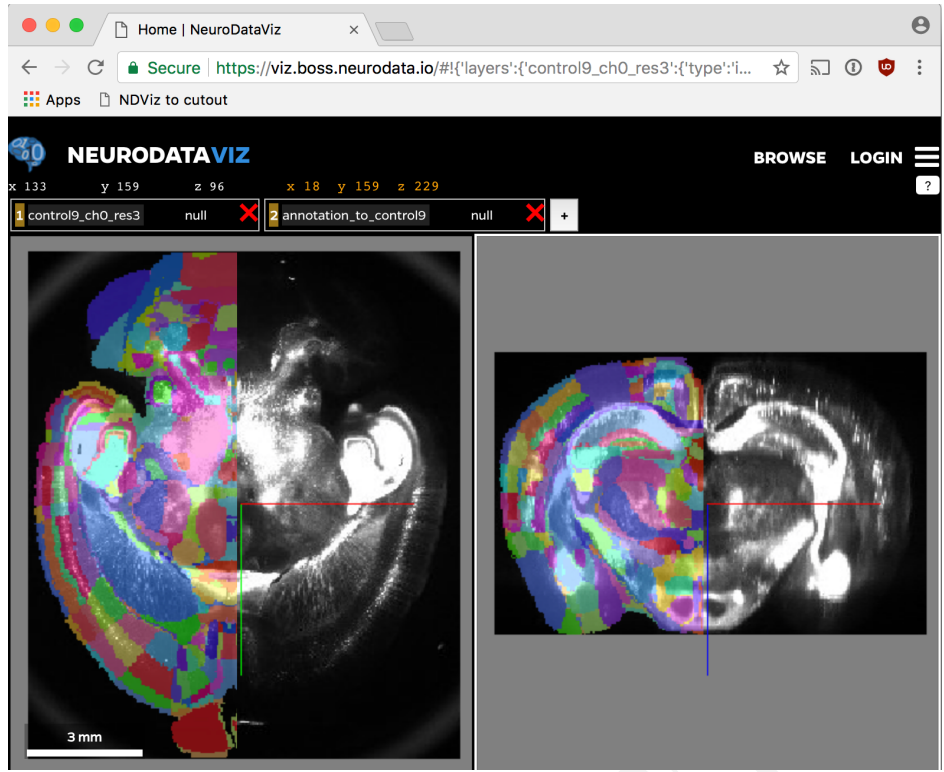

Figure 4: Clarity image with ndreg-registered Allen Institute parcellation overlaid using neuroglancer.

## 3 Conclusion

While this series of workshops have encouraged advances and integration of several platforms and datasets committed to a common vision of globally accessible neuroscience, the bulk of progress must go beyond this work. Forging and maintaining international collaborations is an essential step towards establishing a global brain sciences community that fosters interoperability. We are committed to continuing the support of initiatives that emphasize the development and adoption of platforms, tools, services, standards, and datasets that lower the barrier to performing big data neuroscience reliably and accessibly around the world. We believe that encouraging the interoperability, global democratization, and reproducibility of datasets and analyses is essential for the progress of neuroscience.

## Funding

The hosted events were graciously supported by the National Science Foundation, the Johns Hopkins Kavli Neuroscience Discovery Institute, and the Johns Hopkins Center for Imaging science.

## Notes

[1] http://tinyurl.com/bigbrainboss with username and password *"public"*
[2] https://hub.docker.com/r/bstadt/nomads_deploy/
[3] https://github.com/neurodata/nomads_deploy
[4] https://github.com/neurodata/ingest_large_vol
[5] http://ndwebtools.neurodata.io/
[6] https://github.com/neurodata/ndwebtools
[7] mricloud.org
[8] http://github.com/BIDS-Apps
[9] https://hub.docker.com/r/neurodata/ndreg/
[10] https://github.com/neurodata/ndreg
[11] https://hub.docker.com/r/rymarr/eeg_panda_image/
[12] https://github.com/NeuroDataDesign/orange-panda-f16s17